\documentclass[showpacs,prd,twocolumn,a4paper,amssymb]{revtex4}

\newcommand{\be}{\begin{equation}}
\newcommand{\ee}{\end{equation}}
\newcommand{\beq}{\begin{eqnarray}}
\newcommand{\eeq}{\end{eqnarray}}

\begin{document}
\title{Teleparallel Equivalent of Non-Abelian Kaluza-Klein Theory}
\author{A. L. Barbosa}
\author{L. C. T. Guillen}
\author{J. G. Pereira}
\affiliation{Instituto de F\'{\i}sica Te\'orica,
Universidade Estadual Paulista\\
Rua Pamplona 145,
01405-900 S\~ao Paulo SP, Brazil}
\date{\today}

\begin{abstract}
Based on the equivalence between a gauge theory for the translation group and general
relativity, a teleparallel version of the non-abelian Kaluza-Klein theory is constructed. In
this theory, only the fiber-space turns out to be higher-dimensional, spacetime being kept
always four-dimensional. The resulting model is a gauge theory that unifies, in the
Kaluza-Klein sense, gravitational and gauge fields. In contrast to the ordinary
Kaluza-Klein models, this theory defines a natural length-scale for the compact
sub-manifold of the fiber space, which is shown to be of the order of the Planck length.
\end{abstract}

\maketitle

\section{Introduction}

In ordinary Kaluza-Klein theories~\cite{okk}, the geometrical approach of general relativity
is adopted as the paradigm for the description of all other interactions of nature. In the
original Kaluza-Klein theory, for example, gravitational and electromagnetic fields are
described by a Hilbert-Einstein lagrangian in a five-dimensional spacetime. In the case of the
non-abelian gauge theory, the unification with gravitation, a possibility first raised up in
the sixties~\cite{paper11}, was achieved by extending the usual four-dimensional spacetime to a
($4+D$)-dimensional spacetime~\cite{paper13,paper12,paper15,paper14}, with $D$ the dimension of
the compact part of spacetime. According to this construction, the isometries of the
$D$-dimensional compact sub-manifold yield the non-abelian gauge transformations.

On the other hand, it is widely known that, at least macroscopically, general relativity is
equivalent to a gauge theory~\cite{h} for the translation group~\cite{hn}, provided a
specific choice of parameters be made~\cite{hs}. In this theory, known as the {\it teleparallel
equivalent of general relativity}, the fundamental field is the Weitzenb\"ock connection, a
connection presenting torsion, but no curvature. Differently from general relativity, in which
gravitation is attributed to curvature, teleparallel gravity attributes gravitation to torsion.
Furthermore, whereas in general relativity curvature is used to {\em geometrize} the
gravitational interaction, in teleparallel gravity torsion plays the role of a {\em
gravitational force}~\cite{paper10}. This agrees with the fact that in any gauge theory, the
classical interaction is always described by a force equation.

Now, the equivalence alluded to above opens new perspectives for the study of unified theories.
In fact, instead of using the geometrical description of general relativity, we can adopt the
gauge description as the basic paradigm, and in this way construct what we call the {\it
teleparallel equivalent of Kaluza-Klein} models. According to this approach, both gravitational
and non-abelian gauge fields turn out to be described by a gauge-type lagrangian. This means
that, instead of obtaining the Yang-Mills construction from geometry, as is usually done in
ordinary Kaluza-Klein models, in the teleparallel Kaluza-Klein the geometry (that is,
gravitation) is obtained from a generalized gauge model. By following this approach, a
teleparallel equivalent of the abelian Kaluza-Klein theory has already been
constructed~\cite{tkk}.

By adopting the gauge description as the basic para\-digm, the purpose of this paper will be to
use the equivalence between general relativity and teleparallel gravity to construct the
teleparallel equivalent of a non-abelian Kaluza-Klein theory. In other words, instead of
extending spacetime to higher dimensions, it is the internal (fiber) space that will be
extended to ($4+D$) dimensions, spacetime being kept always four-dimensional. Similarly to
the ordinary non-abelian Kaluza-Klein model, the gauge transformations will be obtained
from the isometries of the fiber space. This construction will be achieved through the
following steps. In Sec.~II, from the analysis of the dynamics of a particle submitted to both
gravitational and Yang-Mills type fields, the {\em unified} gauge potentials and field
strengths are defined. In Sec.~III the corresponding gauge transformations are obtained from
the isometries of the fiber space, and in Sec.~IV, the unified gauge lagrangian is constructed.
The coupling of matter fields with the unified gauge potential is studied in Sec.~V, where the
explicit dependence of all dynamical variables on the internal coordinates is examined.
Finally, in Sec.~VI, the basic properties of the model are discussed. In particular, it is
pointed out that the teleparallel Kaluza-Klein model defines a natural length-scale for the
compact $D$-dimensional sub-manifold of the fiber space, which is found to be of the order of
the Planck length.

\section{Particle Dynamics and Unified Gauge Potentials}

In what follows, the greek alphabet $\mu, \nu, \rho, \dots  = 0, \dots, 3$ will be used to
denote indices related to spacetime. According to the gauge construction, at each point of
spacetime there is a fiber space, which in our case will be a ($4+D$)-dimensional space given
by the direct product $M^4 \otimes B^D$, where $M^4$ is the tangent Minkowski space, and $B^D$
is the group manifold associated to a Yang-Mills symmetry. The first part of the latin alphabet
$a, b, c, \dots = 0, \dots, 3$ will be used to denote indices related to the Minkowski (or
external) part of the fiber, whereas the second part of the latin alphabet $m, n, \dots = 5,
\dots, 4+D$ will be used to denote indices related to the Yang-Mills (or internal) part of the
fiber. The second part of the latin capital alphabet will be used to denote the whole set of
indices of the internal space, which runs through the values $M, N, \dots = 0, \dots, 3, 5,
\dots 4+D$.
The metric of the $(4+D)$-dimensional internal space $M^4 \otimes B^D$ is
\be
\gamma_{M N} = \left(
                  \begin{array}{cc}
                   \eta_{a b} & 0 \\
                   0 & \gamma_{m n} \\
                  \end{array}
               \right) ,
\label{gmn}
\ee
where $\eta_{a b}$ is the metric of $M^4$, which is chosen to be $\eta_{a b} = {\rm diag} \,
(+1, -1, -1, -1)$, and $\gamma_{m n}$ is the metric of the $D$-dimensional space $B^D$. In
general, $B^D$ is a (compact) curved riemannian space, with $\gamma_{m n} = \gamma_{m n} (x^m)$
a function of the coordinates $x^m$ of $B^D$.

As already said, gravitation will enter as a gauge theory for the translation group, whose
action will take place in the Minkowski space $M^4$. As the dimension of the translation group
is the same as that of the Minkowski space, the first part of the latin alphabet $a, b, c,
\dots = 0, \dots, 3$ will also be used to denote the indices related to the translation group.
The intermediary latin alphabet $i, j, k = 5, \dots, 4+I$ will be used to denote indices
related to the Yang-Mills gauge group, with $I$ denoting the number of generators of the
associated Lie algebra. The latin capital letters $A, B, C, \dots = 0, \dots, 3, 5, \dots, 4 +
I$ will be used to denote the whole set of indices related to the group generators. Notice that
the dimension $4+D$ of the fiber does not need to coincide with the dimension $4+I$ of the gauge
group.

We denote by $A^{a}{}_{\mu}$ the gauge potential related to translations, and by
$A^{i}{}_{\mu}$ the Yang-Mills type gauge potentials. According to the gauge description of
interactions, the action integral describing a particle of mass $m$ and gauge charge
$q_{i}$, under the influence of both a gravitational and a gauge field is
\be 
S = \int_{a}^{b}\left[ - m \, c \, d \sigma -
\frac{1}{c} \left(m\, A^{a}{}_{\mu} \, u_{a} +
g \, A^{i}{}_{\mu} \, q_{i} \right) \, dx^{\mu} \right],
\label{pa} 
\ee 
where $d \sigma = (\eta_{ab} dx^a dx^b)^{1/2}$ is the Minkowski invariant interval, $u_{a} = d
x_a/ d \sigma$ is the tangent space four-velocity, and $q_i$ is the Noether charge
related to the internal gauge transformation~\cite{drechsler}. Notice that the mass $m$ appears
as the gravitational coupling constant, whereas the gauge coupling constant is denoted by $g$.
Notice furthermore that we are assuming the weak equivalence principle, and equating the
inertial and gravitational masses.

The equation of motion following from the action (\ref{pa}) is
\beq
e^{a}{}_{\mu} \frac{du_{a}}{ds} &=& \frac{1}{c^2} \, F^{a}{}_{\mu \nu} u_{a} u^{\nu}
+ \frac{g}{m c^{2}} (\partial_{\mu} A^{i}{}_{\nu} - \partial_{\nu} A^{i}{}_{\mu}) \, q_{i}
\, u^{\nu}  \nonumber \\
&-& \frac{g}{m c^{2}} A^{i}{}_{\mu} \frac{d q_{i}}{ds},
\label{eqm1}
\eeq 
where 
\be F^{a}{}_{\mu \nu} = \partial_{\mu} A^{a}{}_{\nu} - \partial_{\nu} A^{a}{}_{\mu}
\label{eqf1}
\ee 
is the gravitational field strength, $ds = (g_{\mu \nu} dx^\mu dx^\nu)^{1/2}$ is the spacetime
invariant interval, $u^{\nu} = dx^{\nu}/ds$ is the spacetime four-velocity, and
$e^{a}{}_{\mu}$ is the tetrad field
\be
e^{a}{}_{\mu} = \partial_{\mu} x^a + c^{-2} A^{a}{}_{\mu}.
\label{tet1}
\ee
We remark that, whereas the Minkowski indices $a, b, c, \dots$ are raised and lowered with
the Minkowski metric $\eta_{a b}$, the spacetime indices $\mu, \nu, \rho, \dots$ are raised and
lowered with the riemannian metric
\be
g_{\mu \nu} = \eta_{a b} \; e^a{}_\mu e^b{}_\nu,
\ee 
which is used to write the spacetime invariant interval $ds$.
 
Considering now that the gauge charge $q_{i}$ satisfies the Wong equation~\cite{wong} 
\be 
\frac{dq_{i}}{ds} + \frac{g}{\hbar c} f_{i j k} A^{j}{}_{\mu} q^{k} u^{\mu} = 0,
\label{we}
\ee
with $f_{i j k}$ the structure constants of the gauge group, the equation of motion
(\ref{eqm1}) can be rewritten in the form
\be
e^{a}{}_{\mu} \frac{du_{a}}{ds} = \frac{1}{c^2} F^{a}{}_{\mu \nu} u_{a} u^{\nu} +
\frac{g}{m c^{2}} F^{i}{}_{\mu \nu} q_{i} u^{\nu},
\label{eqm2}
\ee
where
\be
F^{i}{}_{\mu \nu} = \partial_{\mu} A^{i}{}_{\nu} - \partial_{\nu} A^{i}{}_{\mu} +
\frac{g}{\hbar c} f^{i}{}_{j k} A^{j}{}_{\mu} A^{k}{}_{\nu}
\label{eqf2}
\ee 
is the gauge field strength. In the absence of Yang-Mills field, the equation of motion
(\ref{eqm2}) can be shown to reduce to the geodesic equation of general
relativity~\cite{paper10}.

The gauge structure of teleparallel gravity allows the definition of a {\em unified} gauge
potential ${\mathcal A}^{A}{}_{\mu}$, which is assumed to have the same dimension of the
gravitational potential $A^a{}_\mu$. Consequently, the internal gauge potential $A^i{}_\mu$
must appear multiplied by an appropriate dimensional factor, which we write in the form
\be
{\mathcal A}^{A}{}_{\mu} \equiv  \left( A^{a}{}_{\mu}, \tilde{A}^{i}{}_{\mu} 
\right) = \left( A^{a}{}_{\mu}, \frac{g}{\kappa} A^{i}{}_{\mu} \right),
\label{gp}
\ee 
where $\kappa$ is a parameter to be determined later. Consequently, if we define
a generalized ($4+I$) Noether charge $p_A = m c u_A$, with 
\be 
u_A \equiv \left(u_a, \tilde{u}_{i} \right)
= \left(u_a, \frac{\kappa}{m} q_{i} \right)  
\ee 
a generalized ($4+I$) ``velocity'', the action (\ref{pa}) can be rewritten in the form 
\be 
S=\int_{a}^{b}\left[- m \ c \ d \sigma -
\frac{1}{c^2} {\mathcal A}^{A}{}_{\mu} \, p_{A} \, dx^{\mu} \right] .
\label{pa2}
\ee

In the same way, we can define now a generalized field strength, 
\be
{\mathcal  F}^{A}{}_{\mu \nu} \equiv  \left( F^{a}{}_{\mu \nu}, \tilde{F}^{i}{}_{\mu \nu}
\right) = \left( F^{a}{}_{\mu \nu},
\frac{g}{\kappa} {F}^{i}{}_{\mu \nu} \right).  
\ee
With these definitions, the equation of motion (\ref{eqm2}) assumes the form 
\be
e^{a}{}_{\mu} \frac{du_{a}}{ds} = c^{- 2}
{\mathcal F}^{A}{}_{\mu \nu} \ u_{A} \ u^{\nu}.
\label{(eqm3)}
\ee
This is the {\em unified} --- gravitational plus Yang-Mills --- analog of the electromagnetic
Lorentz force. Its solution determines the trajectory of the particle under the influence of
both gravitational and Yang-Mills fields.

\section{Generalized Gauge Transformations}

A point in the ($4+D$)-dimensional internal space $M^4 \otimes B^D$ will be denoted by $x^M =
(x^a, x^m)$, where $x^a$ are the coordinates of $M^4$, and $x^m$ the coordinates of $B^D$.
A local transformation of these coordinates, which leaves the metric $\gamma_{M N}$ invariant,
can be written in the form
\be
\delta x^M = \delta \alpha^A K_A x^M,
\label{iso}
\ee 
where $\delta \alpha^A = \delta \alpha^A(x^\mu)$ are the infinitesimal parameters, whose
components are written in the form
\be
\delta \alpha^A (x^{\mu}) \equiv (\delta \alpha^a, \delta \tilde{\alpha}^i) =
\left(\delta \alpha^a, \frac{g}{\kappa c^2} \; \delta \alpha^i \right),
\ee
and $K_A$ represent the generators of the transformations. These generators have the form
\be
K_A = K^{N}{}_{A} \partial_{N},
\ee 
where the coefficients $K^{N}{}_{A}$ are the Killing vectors~\cite{weinberg} associated with
the infinitesimal isometries of the internal space $M^{4} {\otimes} B^{D}$. They form a set of
$4 + I$ linearly independent vectors for this space~\cite{okk}, and are given by
\be
K^{N}{}_{A} = \left(
                  \begin{array}{cc}
                   \delta^{b}{}_{a} & 0 \\
                   0 & K^{n}{}_{i} \\
                  \end{array}
              \right),
\ee
with $\delta^{b}{}_{a}$ the Killing vectors of $M^{4}$, and $K^{n}{}_{i}$ the
Killing vectors of $B^{D}$. The generators are, consequently,
\be
K_a = \delta^{b}{}_{a} \partial_b = P_a,
\ee
which are the isometry generators of $M^4$, and
\be
K_i = K^{n}{}_{i} \partial_{n},
\label{kyang}
\ee 
which are the isometry generators of $B^D$. The coordinate transformations, therefore, are
given by
\be
\delta x^a = \delta \alpha^a,
\label{iso1}
\ee
and
\be
\delta x^n = \frac{g}{\kappa c^2} \; \delta \alpha^i \, K^{n}{}_{i} \equiv
\frac{g}{\kappa c^2} \; \delta \xi^n,
\label{iso2}
\ee
with $\delta \xi^n = \delta \alpha^i \, K^{n}{}_{i}$ the transformation parameter in
the Killing basis.

The generators $K_A$ obey the algebra
\be
[K_A, K_B] = f^{C}{}_{A B} K_C,      \label{cr}
\ee
where $f^{A}{}_{B C}$ are the (dimensional) structure constants, whose components are
\be
f^{A}{}_{ B C} = \left\{
\begin{array} {l@{\quad{\rm for}\quad}l}
f^a{}_{b c} = 0, & A, B, C = a, b, c \\
\tilde{f}^{i}{}_{j k} \equiv \chi \; f^{i}{}_{j k}, & A, B, C = i, j, k.
\end{array}
\right. 
\ee
The constant
\be
\chi = \kappa \, \frac{c}{\hbar},
\label{chi}
\ee 
was introduced for dimensional reasons, and in such a way to yield correct forms for the field
strengths and gauge transformations. We have thus the following commutation relations:
\be 
[K_a, K_b] = 0 ,
\ee
and
\be
[K_i, K_j] = \chi \,{f}^{k}{}_{i j} \, K_k .
\label{ks}
\ee

The generalized derivative, covariant under the transformation (\ref{iso}), is
\be
{\mathcal D}_{\mu} = \partial_{\mu} + c^{-2} {\mathcal A}^{A}{}_{\mu} K_A.
\label{cd} 
\ee 
In fact, as a simple computation shows, its commutator gives rise to the generalized field
strength,
\be
[{\mathcal D}_{\mu}, {\mathcal D}_{\nu}] = c^{- 2}
{\mathcal F}^{A}{}_{\mu \nu} K_A,
\ee
where
\be
{\mathcal F}^{A}{}_{\mu \nu} = \partial_{\mu} {\mathcal A}^{A}{}_{\nu} -
\partial_{\nu} {\mathcal A}^{A}{}_{\mu} + c^{- 2} f^{A}{}_{B C}
{\mathcal A}^{B}{}_{\mu} {\mathcal A}^{C}{}_{\nu}.
\label{deff}
\ee
Using the appropriate definitions, this expression is easily seen to yield the correct
expressions for the gravitational and the Yang-Mills field strengths. We notice in passing that
the tetrad field is given by the covariant derivative of $x^a$, the coordinates of the
non-compact part $M^4$ of the fiber space:
\[
e^a{}_\mu = {\mathcal D}_\mu x^a.
\]
This means that this part of the fiber space presents the soldering
property~\cite{livro}.

Now, from the covariance
of ${\mathcal D}_{\mu}$ under the isometric transformation (\ref{iso}), we obtain the gauge
transformation of the generalized potential:
\be
\delta {\mathcal A}^{A}{}_{\mu} = - c^2 {\mathcal D}_{\mu} \delta \alpha^A \equiv
- c^2 \partial_{\mu} \delta \alpha^A -
f^{A}{}_{B C} {\mathcal A}^{B}{}_{\mu} \delta \alpha^C.
\label{ageral}
\ee
For $A = a$, the usual transformation law for the (abelian) gravitational gauge
potential is obtained:
\be
\delta A^{a}{}_{\mu} = - c^2 \partial_{\mu} \delta \alpha^a.
\ee 
For $A = i$, it gives
\be
\delta A^{i}{}_{\mu} = - {\mathcal D}_{\mu} \delta \alpha^i \equiv
 - \partial_{\mu} \delta {\alpha}^i - \frac{g}{\hbar c}
f^{i}{}_{j k} A^{j}{}_{\mu} \delta {\alpha}^k,
\ee
which is the usual non-abelian gauge transformation. When $\delta \alpha^a = 0$,
therefore, the coordinate transformation (\ref{iso}) leads to a pure gauge transformation.

\section{Gauge Lagrangian and Field Equations}

Considering the generalized field strength ${\mathcal F}^{A}{}_{\mu \nu}$, we
can write the unified lagrangian density for the gauge field as
\be
{\cal L} = \frac{e}{16 \pi G} \left(\frac{1}{4} {\cal F}^{A}{}_{\mu \nu}
{\cal F}^{B}{}_{\theta \rho} g^{\mu \theta} N_{A B}{}^{\nu \rho} \right),
\label{unila}
\ee 
where $e = \det(e^{a}{}_{\mu})$. The algebraic indices $A, B, C, \dots$ are raised and lowered
with the Cartan-Killing metric
\be
\eta_{A B} = \left(
\begin{array}{cc}
\eta_{a b} & 0 \\
0 & \eta_{i j} \\
\end{array} \right),
\label{ck}
\ee
whose components related to the translation group coincide with the lorentzian metric
$\eta_{a b}$ of the Minkowski tangent space. The explicit form of the components of $N_{A
B}{}^{\nu
\rho}$ are
\be
N_{a b}{}^{\nu \rho} = {\mathbb A} \; \eta_{a b} e_{c}{}^{\nu} e^{c \rho} +
{\mathbb B} \; e_{a}{}^{\rho} e_{b}{}^{\nu} + {\mathbb C} \; e_{a}{}^{\nu} e_{b}{}^{\rho},
\ee 
with ${\mathbb A}$, ${\mathbb B}$, ${\mathbb C}$ arbitrary parameters~\cite{hs}, which gives
the lagrangian of the gravitational sector, and
\be
N_{i j}{}^{\nu \rho} = \eta_{i j} e_{c}{}^{\nu} e^{c \rho},
\ee
which gives the lagrangian of the gauge sector. The difference in the form of $N_{A
B}{}^{\nu \rho}$ for the different sectors of the theory is directly related to the fact that,
due to the presence of a tetrad field in the gravitational sector, the algebraic and spacetime
indices of this sector are of the same type, and consequently there are additional ways of
contracting the indices. Since there are no tetrads relating the algebraic and the spacetime
indices in the gauge sector, only the usual contraction is present. For the specific choice of
the parameters,
\[
{\mathbb A} = \frac{\mathbb B}{2} = - \frac{\mathbb C}{4} = 1,
\]
teleparallel gravity yields the so called teleparallel equivalent of general relativity. In
this case, the lagrangian (\ref{unila}) becomes
\be
{\cal L} = \frac{c^4 e}{16 \pi G} S^{\rho \mu \nu} T_{\rho \mu \nu} +
\frac{\kappa^{-2} g^2}{16 \pi G} \; \eta_{i j} \left[\frac{e}{4} F^{i}{}_{\mu \nu}
F^{j \mu \nu} \right],
\label{lagranto}
\ee
where
\[
T_{\rho \mu \nu} = c^{-2} e_{a \rho} F^a{}_{\mu \nu} \equiv
{\Gamma}_{\rho \nu \mu} - {\Gamma}_{\rho \mu \nu}
\]
is the torsion of the Weitzenb{\"o}ck connection ${\Gamma}^{\rho}{}_{\mu \nu} = e^{\rho}{}_{a}
\partial_{\nu} e^{a}{}_{\mu}$, and $S^{\rho \mu \nu}$ is the tensor
\beq
S^{\rho \mu \nu} = \frac{1}{4} (T^{\rho \mu \nu} &+& T^{\mu \rho \nu} -
T^{\nu \rho \mu}) \nonumber \\
&-& \frac{1}{2} \left(g^{\rho \nu} T_{\theta}{}^{\mu \theta} -
g^{\rho \mu}T_{\theta}{}^{\nu \theta} \right).
\eeq

To obtain the correct form of the gauge lagrangian, two conditions must be
imposed on the second term of Eq.\ (\ref{lagranto}). The first one is that~\cite{paper14}
\be
\kappa^{2} = \frac{g^2}{16 \pi G} \equiv \frac{g^2}{{\mathcal G}^2}.
\label{kappa}
\ee
The constant $\kappa$, therefore, is simply the relation between the gauge coupling constant
$g$ and the gravitational coupling constant ${\mathcal G}$. The second condition concerns the
relative signs between the gravitational and the Yang-Mills lagrangians. In order to get the
appropriate sign, it is necessary that
\be
\eta_{i j} = - \delta_{i j}.
\ee
Therefore, the Cartan-Killing metric (\ref{ck}) of the unified gauge group becomes
\be
\eta_{A B} = \left(
                  \begin{array}{cc}
                   \eta_{a b} & 0 \\
                   0 & - \delta_{i j} \\
                  \end{array}
               \right) .
\label{gck}
\ee
With these conditions, we obtain
\be 
{\cal L} \equiv {\cal L}_{\rm G} + {\cal L}_{\rm YM} =
\frac{c^4 e}{16 \pi G} S^{\rho \mu \nu} T_{\rho \mu \nu} - \frac{e}{4}
F^{i}{}_{\mu \nu} F_{i}{}^{\mu \nu}.
\label{lagf}
\ee
As is well known, up to a divergence, the first term of this lagrangian is the teleparallel
equivalent of the Einstein-Hilbert lagrangian of general relativity~\cite{paper10}. The second
term, on the other hand, is the usual gauge lagrangian in the presence of gravitation.

It is interesting to notice that, when the Cartan-Killing metric related to the (external)
translation group $T_4$ is chosen to be $\eta_{ab} = {\rm diag} \, (+1, -1, -1, -1)$, the
corresponding Cartan-Killing metric related to the (internal) Yang-Mills group $G_{\rm
YM}$ has necessarily the form $\eta_{ij} = - \delta_{ij}$. If we had chosen the other possible
convention for $\eta_{ab}$, that is, $\eta_{ab} = {\rm diag} \, (-1, +1, +1, +1)$, the
consistency condition would require that $\eta_{ij} = + \delta_{ij}$. Therefore, the
teleparallel Kaluza-Klein construction imposes a constraint between the Cartan-Killing metric
convention adopted for the translation gauge group, and consequently for the Minkowski tangent
space --- see the comment just after Eq.\ (\ref{ck}) --- and that adopted for the Yang-Mills
gauge group.

Performing a functional variation of ${\cal L}$ in relation to the components $A^{a}{}_{\tau}$,
and using the definition of the Weitzenb{\"o}ck connection ${\Gamma}^{\rho}{}_{\mu \nu}$, we
obtain the gravitational field equation 
\be
\partial_{\sigma} (e S_{\lambda}{}^{\tau \sigma}) - \frac{4 \pi G}{c^4} \, e \,
t^{\tau}{}_{\lambda} = \frac{4 \pi G}{c^4} \, e \, {\Theta}^{\tau}{}_{\lambda},
\label{eqs}
\ee 
where
\be
e \, t^{\tau}{}_{\lambda} = \frac{c^{4} e}{4 \pi G} \; {\Gamma}^{\mu}{}_{\nu 
\lambda} \; S_{\mu}{}^{\tau \nu} - \delta ^{\tau}{}_{\lambda} \, {\cal L}_{\rm G}
\ee
stands for the teleparallel canonical energy-momentum pseudo-tensor of the gravitational
field~\cite{prl}. The source of the field equation (\ref{eqs}) is the energy-momentum tensor of
the Yang-Mills field in the presence of gravitation, which is defined by
\be
e {\Theta}^{\tau}{}_{\lambda} \equiv - c^{-2} e^{a}{}_{\lambda} \frac{\delta {\cal
L}_{\rm YM}}{\delta A^{a}{}_{\tau}} = - e^{a}{}_{\lambda} \frac{\delta {\cal
L}_{\rm YM}}{\delta e^{a}{}_{\tau}}.
\label{tetaym}
\ee
Finally, variation of ${\cal L}$ with respect to the components $A^{i}{}_{\mu}$ yields the
Yang-Mills equation in the presence of gravitation,
\[
\partial_\mu( e F^{i \mu \nu}) -
\frac{g e}{\hbar c} \; j^{i \nu} = 0,
\]
where
\[
j^{i \nu} = - f^i{}_{j k} \; A^j{}_\mu \; F^{k \mu \nu}
\]
stands for the (pseudo) current of the Yang-Mills field \cite{ramond}. We mention in passing
that the teleparallel field equation (\ref{eqs}) is an equation written in terms of the
Weitzenb\"ock connection only. It can alternatively be written in terms of the Levi-Civita
connection, in which case it reduces to the general relativity Einstein's equation. The
teleparallel field equation (\ref{eqs}), however, has the advantage of presenting the same
formal structure of the Yang-Mills equation.

\section{Matter Fields}

The geometrical structure underlying every gauge theory exists independently of the presence or
not of gauge fields. For example, in the teleparallel Kaluza-Klein theory, the non-compact
four-dimensional part of the fiber, which is given by the tangent Minkowski space, is a
geometrical structure that is always present independently of the presence or not of a
gravitational gauge field. The same is true of the $D$-dimensional compact part of the fiber in
relation to the corresponding Yang-Mills gauge field. It should be noticed that the fiber
space of teleparallel Kaluza-Klein theories corresponds to the ground state spacetime of
ordinary Kaluza-Klein theories. In these theories, the transition from a higher-dimensional
theory to the effective four-dimensional theory is made with the help of an harmonic expansion
around the ground state, whose excitations represent the field variables of the model. As a
consequence, an infinite spectrum of particles is obtained. In particular, the lowest order
excitations have vanishing mass, giving rise to the massless sector of the emerging gauge
theory.

On the other hand, in the teleparallel Kaluza-Klein theories, all dynamical variables are
function of the four-dimensional spacetime points. Furthermore, the action functional and the
field equations are written in the four-dimensional spacetime, and not in the higher-dimensional
fiber space. This means that no dimensional reduction is necessary, no harmonic expansion
around the ground state has to be performed, and consequently the field variables of the model
cannot be represented by excitations. In fact, like in any other gauge theory, the basic
fields are represented by gauge potentials given {\it a priori}, which are the basic
ingredients for the construction of gauge theories.

Now, comes the question on how the dynamical variables of the theory depend on the coordinates
$x^M$ of the fiber space. Concerning the coordinates $x^a$ of the non-compact four-dimensional
part ($M^4$) of the fiber, as this space is soldered to spacetime, and as all dynamical
variables are functions of the spacetime coordinates $x^\mu$, the dependence of these variables
on $x^a$ has necessarily to be through the argument of the dynamical variables. Concerning the
coordinates $x^m$ of the compact $D$-dimensional part ($B^D$) of the fiber, as a change in
$x^m$ must correspond to a change, not in the argument, but in the components of the field
variable, the dependence of any dynamical variable on $x^m$ will be analogous to the
dependence on the gauge parameter in a gauge theory. This is a direct consequence of the fact
that the isometry transformations of $B^D$ are ultimately equivalent to internal gauge
transformations. Accordingly, the dependence of the matter field $\Psi$ on the coordinates
$x^m$ can be written in the form
\be
\Psi(x^m) = \exp \left[ i \;\chi \; \beta_n x^n \right] \psi,
\label{psi2}
\ee 
where $\chi$ is defined in Eq.\ (\ref{chi}), and $\beta_n$ are parameters related to the
geometry of the compact manifold $B^D$. In addition, as a change in $x^m$ is related to a gauge
transformation, $\beta_n$ must necessarily assume values in the Lie algebra of the gauge group.
In other words, $\beta_n = \beta_n{}^j \, {\mathcal T}_j$, with ${\mathcal T}_j$ a matrix
representation of the Lie algebra generators. In fact, according to Eq.\ (\ref{psi2}), the
action of the (derivative) isometry generators $K^n{}_i \partial_n$, turns out to be equivalent
to the action of the (multiplicative) matrix generators $i \chi K^n{}_i \beta_n$. This means
that it is possible to relate $i \chi K^n{}_i \beta_n$ to another realization of the generators
of the gauge group. As already said, this possibility is a direct consequence of the fact that
the (internal) gauge transformations are obtained as the isometries of
$B^D$. 

Let us explore better this point. Under the coordinate transformation (\ref{iso}), a matter
field $\Psi$ changes according to
\be 
\delta \Psi =
\delta \alpha^A K_A \Psi \equiv \delta \alpha^a \partial_a \Psi +
\frac{g}{\kappa c^2} \; \delta \alpha^i K^{n}{}_{i} \partial_{n} \Psi.
\label{var}
\ee 
By using Eq.\ (\ref{psi2}), we see that
\be
\partial_{n} \Psi = i \, \chi \, \beta_n \Psi =
i \; \frac{\kappa c}{\hbar} \, \beta_n \Psi,
\label{dp}
\ee
where use has been made of Eq.\ (\ref{chi}). Substituting into the transformation (\ref{var}),
it becomes 
\be 
\delta \Psi =
\delta \alpha^a \, \partial_a \Psi +
\frac{i g}{\hbar c} \; \delta \alpha^i K^{n}{}_{i} \beta_{n} \Psi.
\label{var2}
\ee 
On the other hand, we have already seen that $K_i = K^{n}{}_{i} \partial_{n}$ satisfy the
commutation relations (\ref{ks}), that is
\[
[K^{n}{}_{i} \partial_{n}, K^{m}{}_{j} \partial_{m}] \Psi =
\chi \, f^k{}_{i j} \, K^{n}{}_{k} \partial_{n} \Psi.
\]
As the Killing vectors $K^n{}_i$ depend on $x^m$ in the same manner as $\Psi$ does, it is an
easy task to verify that
\be
[i K^{n}{}_{i} \beta_{n}, i K^{m}{}_{j} \beta_{m}] \Psi =
f^k{}_{i j} \, i K^{n}{}_{k} \beta_{n} \Psi.
\label{ula}
\ee
This means that
\be
T_i = i K^{n}{}_{i} \beta_{n} \equiv i K^{n}{}_{i} \, \beta_n{}^j \, {\mathcal T}_j
\ee
can be identified as another realization of the (anti-hermitian) Lie algebra generators.
With this identification, the transformation (\ref{var2}) acquires the form
\be 
\delta \Psi = \delta
\alpha^a P_a \Psi + \frac{g}{\hbar c} \delta {\alpha}^i \, T_{i} \, \Psi,
\label{var3}
\ee
which is in fact a gauge transformation of matter fields.

The covariant derivative of $\Psi$ is defined by
\be
{\mathcal D}_{\mu} \Psi = \partial_\mu \Psi + {\mathcal A}^A{}_\mu \,
\frac{\delta \Psi}{\delta \alpha^A}.
\ee
Substituting the transformation (\ref{ageral}), and using the appropriate identifications, it
becomes
\be 
{\mathcal D}_{\mu} \Psi =
e^{a}{}_{\mu} \partial_{a}\Psi + 
\frac{g}{\hbar c} A^{i}{}_{\mu} \, T_i \, \Psi,
\label{dev3}
\ee
which is the usual expression of the gauge covariant derivative in the presence of
gravitation. Defining $A^{i}{}_{\mu} = A^{i}{}_a \, e^{a}{}_{\mu}$, it can be rewritten in the
form
\be
{\mathcal D}_{\mu} \Psi =
e^{a}{}_{\mu} \, {\mathcal D}_a \Psi,
\ee
where ${\mathcal D}_a \Psi$ is the gauge covariant derivative in Min\-kowski spacetime.

\section{Final Remarks}

Replacing the general relativity par\-a\-digm by a gauge par\-a\-digm, and making use of the
teleparallel description of gravitation, which corresponds to a gauge theory for the
translation group, we have succeeded in constructing a teleparallel version of the non-abelian
Kaluza-Klein theory. In other words, we have succeeded in unifying, in the Kaluza-Klein sense,
teleparallel gravitation with Yang-Mills type theories. The resulting model turns out to be a
gauge theory for the group $T_4 \otimes G_{\rm YM}$, with the fiber space given by $M^4 \otimes
B^D$, where $M^4$ is the Minkowski tangent spacetime, and $B^D$ is the manifold associated to
the Yang-Mills gauge group $G_{\rm YM}$. In this model, the translational gauge transformation
arises as the isometries of the non-compact four-dimensional part of the fiber, which is always
a Minkowski spacetime $M^4$, whereas the non-abelian gauge transformations arise as the
isometries of the compact $D$-dimensional part of the fiber, which is the part related to the
internal gauge symmetry.

As in the abelian case~\cite{tkk}, the teleparallel equivalent of the non-abelian Kaluza-Klein
model turns out to be much more natural than ordinary's Kaluza-Klein. In fact, in the
teleparallel model both gravitational and Yang-Mills type fields are described by a gauge
theory, with the Yang-Mills field-strength appearing as extra gauge-components of torsion, the
field strength of teleparallel gravity. This means that the gravitational and the Yang-Mills
field-strengths are different components of a unique tensor. Another interesting point concerns
the relation between geometry and gauge theories. According to ordinary Kaluza-Klein models,
gauge theories emerge from higher-dimensional geometric theories as a consequence of the
dimensional reduction process. According to the teleparallel Kaluza-Klein approach, however,
gauge theories are the natural structures to be introduced, the four-dimensional geometry
(gravitation) emerging from the non-compact sector of the fiber space. In fact, only this
sector presents the soldering property~\cite{livro}, and can consequently give rise to a tetrad
field, which is the responsible for the geometrical structure (either metric or teleparallel)
induced in spacetime. Furthermore, as the gauge theories are introduced in their original form
--- they do not come from geometry --- the unification, though not trivial, turns out to be
much more natural and easier to be performed.

An important characteristic of the ordinary non-abe\-lian Kaluza-Klein model is that the metric
(\ref{gmn}) is not a solution of the higher-dimensional Einstein equations as these equations
cannot have solutions of the form $M^4 \otimes B^D$. This is related to the fact that the
$D$-dimensional internal space is in general curved, leading then to difficulties for defining
the ground state (vacuum) of the higher-dimensional gravitational field. These models,
therefore, require an initial non-compact $(4+D)$-dimensional spacetime, and a subsequent
compactification scheme for the $D$ extra dimensions. One way of solving this problem is to
introduce extra matter fields in the form of a higher-dimensional energy-momentum tensor, so
that an spontaneous compactification of the extra dimensions is achieved~\cite{cremmer}. Another
solution was that provided by Freund and Rubin~\cite{freru} in eleven-dimensional supergravity,
where not only there is compactification, but the space naturally separates in $(4+7)$
dimensions. On the other hand, since in the teleparallel Kaluza-Klein model the gauge theories
are not obtained from the geometry, but introduced in their original forms, the fiber space of
these theories can be assumed to present a compact sub-manifold $B^D$ from the very beginning.
In other words, the compactification problem does not exist for these theories. In addition, as
both the action and the field equations are always written in the four-dimensional spacetime,
and not in the higher-dimensional fiber space, no dimensional reduction is necessary, and
consequently no expansion of the dynamical variables in terms of the complete set of harmonics
of $B^D$ has to be performed. As a consequence, the infinite spectrum of new particles
is absent, strongly reducing the redundancy present in ordinary Kaluza-Klein theories. A
similar achievement has already been obtained by a modified Kaluza-Klein theory in which the
internal coordinates are replaced by generators of a non-commutative algebra~\cite{madore}. In
this model, no truncation to eliminate extraneous modes is necessary as only a finite number of
them is present.

Finally, as a last remark, let us take the internal coordinate transformation
(\ref{iso2}), and substitute on it the value of $\kappa$, given by Eq.\ (\ref{kappa}). As $B^D$
is compact, this transformation can be written in the form
\be
\delta x^n = \rho \; \delta \theta^n,
\ee
where
\be
\rho = l_P \equiv \left(\frac{G \hbar}{c^3} \right)^{\frac{1}{2}},
\ee
is the length-scale associated to the compact internal sub-manifold, and
\be
\delta \theta^n = \left(\frac{16 \pi}{\hbar c} \right)^\frac{1}{2} \delta \xi^n
\ee
is the angular coordinate associated to $x^m$. We see in this way that the teleparallel
Kaluza-Klein model defines a natural length-scale for the compact part of the fiber space,
given by the Planck length. In the specific case of the ordinary abelian Kaluza-Klein theory,
the radius of the fifth dimension can only be inferred from the value of the elementary
electric charge. Since the teleparallel model yields a natural length-scale, we can reverse
the argument and use this length to {\em calculate} the elementary electric charge.
Furthermore, as is well known, a length of the order of the Planck length, like the $\rho$
above, through the application of the Bohr-Sommerfeld quantization rule to the periodic motion
in the fifth dimension, gives the correct value for the elementary electric charge.

\section*{Acknowledgments}

The authors would like to thank FAPESP-Brazil, CAPES-Brazil and 
CNPq-Brazil for financial support.

\end{document}